
\magnification=\magstep1

\input psfig

\def \lp {\left(}
\def \rp {\right)}
\def\barp{{\bar p}}
\def\baru{{\bar u}}

\def\barz{{\bar z}}
\def\barW{{\bar W}}
\def\vare{\varepsilon}
\def\nid{\noindent}
\font\title=cmbx8 scaled \magstep2
\overfullrule=0pt

\centerline{\title A Generalization of the
Theory of Normal Forms}

\medskip

\centerline{W. H. Warner, P. R. Sethna}
\centerline{Aerospace Engineering and Mechanics}
\centerline{University of Minnesota}
\centerline{and}
\centerline{James P. Sethna}
\centerline{Physics}
\centerline{Cornell University}
\centerline{chao-dyn/9510014}
\baselineskip 20pt
\bigskip

Normal form theory is a technique for transforming the ordinary differential
equations describing nonlinear dynamical systems into certain standard forms.
Using a particular class of coordinate transformations, one can remove the
inessential
part of higher-order nonlinearities.  Unlike the closely-related method of
averaging,
the standard development of normal form theory involves several technical
assumptions
about the allowed classes of coordinate transformations (often restricted to
homogeneous
polynomials).  In a recent paper [1], the second author considered the
equivalence of the
methods of averaging and of normal forms.  The references given there,
particularly Chow and Hale [2], should be consulted for a full treatment of Lie
Transforms.

In this paper, we relax the restrictions on the transformations
allowed.
We start with the Duffing equation, and show that a singular coordinate
transformation
can $\it {remove}$  the nonlinearity associated with the usual normal form.  We
give two
interpretations of this coordinate transformation, one with a branch cut
reminiscent of a Poincar{\'{e}} section.  We then show, when the generating
problem is linear and autonomous with diagonal Jordan form, that we can
remove all nonlinearities order by
order using singular coordinate transformations generated  by the solution to
the first-order linear partial differential equation produced by the Lie
Transform method of normal form theory. A companion paper [4] discusses these
methods in a
more general context and treats a specific example with a nondiagonal Jordan
form for the
generating matrix.

\vfill\eject

\nid {\bf 1. Duffing's Equation: Removing the Nonlinearities}

The second-order problems modeled by Duffing's equation or
 van der Pol's equation under near-resonance forcing are examples of
 the behavior that concerns us here.
 We use the notation of the Duffing example in [1], Eqn. (37) :
$$
{\ddot x}+ \omega^2x + \varepsilon{\tilde c}{\dot x} + \varepsilon{\tilde h}
 x^3 = \vare {\tilde R} \cos \lp\Omega t\rp
\eqno (1)
$$
where $\Omega \approx \omega, \varepsilon$ is a small parameter, and the other
constants are positive. If we manipulate the variables and parameters in a
standard way (see[1]) by choosing
$$
{\tilde R}=\Omega^2 R,~{\tilde h}= \Omega^2 h,~{\tilde c}=\Omega c,~
\omega^2 = \Omega^2 \lp 1+ \varepsilon\sigma \rp,~ \tau = \Omega t, ~y = dx /
d\tau = x^\prime
$$
and changing to complex state variables
$$
p \lp\tau\rp = {1\over 2} \lp x+ iy \rp ,\dots {\barp} ={1 \over 2} \lp x-iy
\rp,
$$
then  Eqn. (1) becomes
$$
p^{\prime}= -ip+i\varepsilon\left\lbrace-{1\over 2}(\sigma - ic)p-{1\over 2}
(\sigma + ic){\bar p}-{h\over 2}(p+{\bar p})^3+{R\over
4}(e^{i\tau}+e^{-i\tau})\right\rbrace
\eqno(2)
$$
and its conjugate.

In [1], standard methods are used to show that the coordinate transformation
$$
p=z+ \varepsilon W \lp z,\barz, \tau\rp,\dots \barp = \barz+ \varepsilon\barW
\lp z,
\barz, \tau\rp
\eqno (3)
$$
with
$$
W=-{1 \over4}(\sigma+i c) \barz + {R \over8} e^{i\tau}
	+{h\over4} \lp z^3-3z\barz^2 - {\barz^3 \over 2}\rp
\eqno (4)
$$
can transform the p-equation into the normal form
$$
{z^\prime} = -iz + {\vare f_1} \lp z, \barz , \tau\rp + O \lp \vare^2 \rp
\eqno (5)
$$
where
$$
\eqalign {
f_1 &= K_1z + K_2 z^2 \barz + K_3 e^{-i\tau} ,\cr
K_1 &=-{i \over 2} \lp\sigma -ic \rp,~ K_2 = -{3ih \over 2},~ K_3 = {iR \over
4}}
\eqno (6)
$$
It is the
$z^2 \barz$
term that standard normal form theory considers as essential to keep at the
cubic order for
equations of this type.

Now consider the coordinate transformation from $(p, \barp )$ to $(u, \baru )$
given by
$$
\eqalign
{p &=u+\vare V (u, \baru, \tau)\cr
&=u+ \vare \left[ W\lp u,\baru , \tau\rp +{i\over2} f_1 {\lp u, \baru ,
\tau\rp}
{\rm ~ln} \lp {u \over\baru } \rp \right]}
\eqno (7)
$$
and its conjugate. Explicit computation shows that
$u \lp\tau\rp $
satisfies a differential equation of the form
$$
u^{\prime} =-iu+O\lp\vare^2\rp .
\eqno (8)
$$
By using a coordinate transformation with a logarithmic singularity at
$u=0$, we have removed the first-order term in $ \varepsilon $,
including the $ u^2 \baru $ term which normal form theory tells us is
essential and irremovable.

A special case of Equation (5) occurs when
$f_1 = z - z^2\barz $ (note that this does not correspond to a Duffing equation
with real
coefficients).  For this special case, the exact solution
$ z(\tau) = r(\tau) $ exp $(i\theta (\tau))$ to the nonlinear equation through
order $\vare $
terms can be written down for initial conditions
$ r(0) = r_0 \neq 1, \theta (0) = \theta_0:$
$$
r(\tau ) ={r_0 \over {{\sqrt {{r^2_0} + (1-{r^2_0})~ \exp [-2\vare \tau]}}}},~
 {\theta (\tau) = +{\theta_0} -\tau .} $$
Our transformation  Equation (7) generates the approximate solution
$$ r^{(1)} (\tau )  = r_0 + r_0 (1- {r_0^2}) (\vare \tau ),~ \theta^{(1)} =
\theta (\tau )
= \theta_0 - \tau
$$
Graphs of both types of functions starting from the same initial conditions
$z(0) = (1 + i)/2$
are shown in Figures 1 and 2.

\vfill\break
\null\vfill

\centerline{
\psfig{figure=Duffing.ps,width=4truein}
}

Figure 1 shows the trajectory corresponding to the exact solution $r(\tau)$ for
$\vare=0.05$.

\vfill\null
\break
\null\vfill

\centerline{
\psfig{figure=Duffing_bill.ps,width=4truein}
}
Figure 2 shows the corresponding graph of the
approximate solution $r^{(1)}(\tau)$. It is
clear that the approximate solution spirals outward, crossing the limit
cycle after a finite time $ \tau_0$ of order $1/\vare$. One can verify
that our approximate solution is the first term in the power series
expansion of the exact solution in the parameter $\vare $.  Notice
that the approximation is good to first order in $\vare$, but not uniformly
accurate in time.

\vfill\null
\break
\null\vfill

\centerline{
\psfig{figure=Duffing_poincare.ps,width=4truein}
}

Figure 3 shows trajectories for various initial conditions generated by
our coordinate transformation $V$.  Here we have restricted $V$ to a
single Riemann sheet of the logarithm: hence the discontinuity at the
branch cut along $Re[z]<0$, $Im[z]=0$.  This branch cut is what allows
the logarithm to ``unwrap'' the singularity of the Duffing equation.
Thus it is natural {\it not} to continue increasing $\theta$ with time,
but to restart the approximate solution every time one crosses the
negative real axis.  This may at first seem strange, as we are making a jump in
the
nonphysical variable $u(\tau)$.  On the other hand, we are forced into this to
make the
coordinate transformation $V(u, \baru, \tau)$ a single-valued function: the
discontinuity in
$u$ is needed to make the dynamics of $p=u+\epsilon V$ continuous.
Our $O(\varepsilon)$ solution now systematically approximates a
{\it Poincar\'e first-return map} $T$ along the negative real
axis:  $$T^{(1)}(r) = r + r(1-r^2)(2 \pi \varepsilon)$$
giving a systematic approximation for the next intersection $T(r)$ of a curve
which crosses $Im[z]=0$ at $r = Re(z) < 0$.  This interpretation of the
dynamics preserves the qualitative behavior of the original dynamical system,
although admittedly does not produce an explicit analytical solution for
the dynamics.

\vskip 36pt
\nid {\bf 2. Solving the Lie Differential Equation}

Why does normal form theory miss this useful transformation?
(Alternatively, how does normal form theory avoid this nasty singular
transformation?) We must look more closely at how the transformation of
state variables is found by solving the partial differential equation
produced by the Lie transform process.

 For the case when the state equations have linear generating terms in diagonal
  Jordan form
$({\bf A} = {\rm diag} (\lambda_\alpha) ~{\rm {\hbox{with no}}}~
\lambda_\alpha =0, \alpha =1,2,\dots, N)$
, the state equation governing the component $x_\alpha$ of ${\bf {x}}$ will be
$$
{dx_\alpha \over dt}= \lambda_\alpha x_\alpha + \vare {f^{(1)}_\alpha} (t,{\bf
{\hbox {x}}}) + {\vare^2 \over 2} {f^{(2)}_\alpha} (f, {\bf {\hbox {x}}}) +
\dots
$$
The equation for the corresponding component ${W^{(1)}_\alpha } \equiv W_\alpha
 (t, {\bf{\hbox {y}}})$ in the transformation $ {\bf {\hbox {x}}} =
 {\bf {\hbox {y}}} + \vare W^{(1)} + (\vare^2/2)W^{(2)} +
 \dots$ is the Lie equation
$$
L(W_\alpha) \equiv {\partial W_\alpha \over\partial t} + \sum_\beta
\lambda_{_{\beta}}
y_{_{\beta }}{\partial W_\alpha \over \partial y_{_{\beta}}} -\lambda_\alpha
W_\alpha
=f^{(1)}_\alpha (t, {\bf {\hbox {y}}}).
$$
For autonomous ${{\bf {\it f}}}^{(1)}$, the $\partial/\partial t$ term on the
left is dropped
from $L$. In that case, by restricting the class of coordinate transformations
to homogeneous
polynomials, normal form theory cannot remove any nonlinearities in ${{\bf {\it
f}}}^{(1)}$
 that lie in the null space of the Lie operator .  These nonlinear terms
therefore comprise
the ``normal form": all other terms are removed by the coordinate
transformation.
But why should the null space of $L$ (the set of functions satisfying $L(W)=0)$
have
anything to do with finding a particular solution to the equation $L(W)=f$?

To understand this, consider a different linear operator, corresponding to
solutions
 of the forced harmonic oscillator:
$$
H (x) \equiv {\ddot{x}} +x=f (t)
$$
There is something special about solutions $x(t)$ for forcing functions in the
null
space of $H$.  If we force at resonance $f(t) = \sin (t)$, then ${\it f}$ is in
the null
space of $H : H(f)=0,$ and we see that the particular solution $x(t) = t ~\sin
(t)$ is
 qualitatively different from the solution for $f(t) = \sin (\omega t)$ for
other frequencies $\omega$.
 If we restrict the class of perturbations and solutions to finite sums of
harmonic waves,
 then there would be no solution to $H(f) = \sin (t)$.
 Normal form theory makes precisely this kind of restriction: by restricting
the
 perturbations and solutions to be homogeneous polynomials or in the
Hamiltonian
case
 to canonical transformations, they have defined away the possibly singular
 coordinate transformations that we study here.

We now demonstrate, for perturbations of the form
${{\bf f}^{(1)}_\alpha} (t, {\bf {\hbox {y}}}) $
above, that we can find solutions W to the Lie operator partial differential
equation
$$
L(W_\alpha) \equiv {\partial W_\alpha \over\partial t} + \sum_\beta
\lambda_{_{\beta}}
y_{_{\beta }} {\partial W_\alpha
\over \partial y_\beta} -\lambda_\alpha W_\alpha =f^{(1)}_\alpha (t, {\bf
{\hbox
{y}}}).
$$
in complete generality by reducing it to an ordinary one.  Using the method of
characteristics, one discovers (see Courant-Hilbert [3], p. 11, for a related
transformation)
that the coordinate transformation  from $({\bf {\hbox {y}}},t)$ to $(\xi,
\tau)$ given by
$$
{\eqalign {\xi_1 &=y_1;\cr
\xi_\beta &=(y_\beta^{1/\lambda_\beta}) /(y_1^{1/ \lambda_1}),~~~~~ \beta
=2,3,\dots, N;\cr
\tau &= y_1 \exp(-\lambda_1 t) \cr}}
$$
will reduce the partial differential equation to the ordinary differential
equation
$$
\lambda_1 \xi_1 {\partial V \alpha \over {\partial\xi_1}}= G_\alpha (\xi_1 ,
\xi_2 , \dots ,
\xi_n ,\tau)={1 \over {y_\alpha}} {f^{(1)}_\alpha} (t, {\bf {\hbox {y}}}).
$$
where the $V_\alpha = W_\alpha /y_\alpha$ are now to be considered as functions
of
$(\xi ,\tau)$ and the $G$'s are equal to the functions on the far right
evaluated
in the new variables. Our solution (7) removing the O($\varepsilon$) term in
the Duffing equation, was generated using precisely this method.

\vskip 36pt
\nid {\bf 3. Concluding Remarks}

We have studied the Duffing equation using a new approach to the calculation
of an approximate solution.  What about our methods in general?

Our use of the method of characteristics is perfectly general:  the same
coordinate transformations used in the theory of normal forms can be shown [4]
to remove all nonlinearities if the space of allowed functions is not
restricted.  Important questions remain about the nature of the higher order
terms in $\vare$, and about the estimates of the finite time for which the
approximate solution is valid.  (Here we find results valid to times of order
$1/\vare$, but for the example of [4] times of order $1/\vare^{(1/4)}$ are
found.)

Our simple interpretation of the resulting logarithmic transformation as a
Poincar\'{e} return map gave us a correct qualitative picture of the
global dynamics for the Duffing equation. We do not have a general formula
for analyzing other systems in this way, but we find intriguing the implied
link between the singularities of the coordinate transformations
introduced by the method of characteristics and the qualitative structure
of the corresponding dynamics.

\vfill\eject
\nid {\bf References}
\baselineskip 12pt
\parskip 22pt

\item {1.}  P.R. Sethna, {\it On Averaged and Normal Form Equations}, Nonlinear
Dynamics, {\bf 7},
1-10 (1995).

\item {2.}  S.N. Chow and J. R. Hale, {\it Methods of Bifurcation Theory},
Springer-Verlag, New York
(1982).

\item {3.}  R. Courant and D. Hilbert, {\it Methods of Mathematical Physics},
Vol. II: Partial
Differential Equations, Wiley: Interscience Publishers, New York (1962).

\item {4.}  W. H. Warner, {\it An Extension of Normal Form Methods for
Calculating
Approximate Solutions}, submitted for publication (1995).

\end